\documentclass[twocolumn,showpacs,preprintnumbers,amsmath,amssymb,aps,prb,10pt]{revtex4-1}

\usepackage{graphicx}
\usepackage{subfigure}
\usepackage{dcolumn}
\usepackage{bm}

\newcommand{\beq}{\begin{eqnarray}}
\newcommand{\eeq}{\end{eqnarray}}

\newcommand{\bfig}{ \left. \begin{array}{l} }
\newcommand{\bfigor}{ \left\{ \begin{array}{ll} }
\newcommand{\efig}{ \end{array} \right. }

\newcommand{\lan}{\langle}
\newcommand{\ran}{\rangle}

\newcommand{\al}{\alpha}
\newcommand{\be}{\beta}

\newcommand{\cH}{{\cal H}}

\newcommand{\cF}{{\cal F}}
\newcommand{\bS}{{\bf S}}
\newcommand{\tilh}{\tilde{h}}

\newcommand{\non}{\nonumber}

\newcommand{\la}{\lambda}

\newcommand{\w}{\omega}

\newcommand{\e}{\epsilon}

\newcommand{\Ga}{\Gamma}

\newcommand{\krest}{\dagger}
\newcommand{\akrest}{a^\krest}

\newcommand{\akr}{\akrest}

\newcommand{\kk}{{\bf k}}
\newcommand{\qq}{{\bf q}}
\newcommand{\pp}{{\bf p}}

\begin{document}

\title{Spin nematic states in spin-1 antiferromagnets with easy-axis anisotropy}

\author{A.\ V.\ Sizanov$^1$}
\email{alexey.sizanov@gmail.com}
\author{A.\ V.\ Syromyatnikov$^{1,2}$}
\email{syromyat@thd.pnpi.spb.ru}
\affiliation{$^1$Petersburg Nuclear Physics Institute NRC "Kurchatov Institute", Gatchina, St.\ Petersburg 188300, Russia}
\affiliation{$^2$Department of Physics, St.\ Petersburg State University, 198504 St.\ Petersburg, Russia}

\date{\today}

\begin{abstract}

It is well known that spin nematic phases can appear in either frustrated magnets or in those described by Hamiltonians with large exotic non-Heisenberg terms like biquadratic exchange. We show in the present study that {\it non-frustrated} spin-1 1D, 2D and 3D antiferromagnets with single-ion easy-axis anisotropy can show nematic phases in strong magnetic field. For 1D case we support our analytical results by numerical ones.

\end{abstract}


\maketitle

{\bf 1. Introduction}. 
Frustrated spin systems have offered in recent years a wealth of opportunities for the study of a broad range of novel types of states and phase transitions. Spin nematic phases form a class of objects in this area which has received much attention. Spin nematic states are spin-liquid-like states which show a multiple-spin ordering without the conventional long-range magnetic order. The two-spin ordering can be generally described by the tensor \cite{andreev} $Q_{jl}^{\alpha\beta} = \langle S_j^\alpha S_l^\beta\rangle-\delta_{\alpha\beta} \langle {\bf S}_j {\bf S}_l \rangle /3$. The symmetric part of $Q_{jl}^{\alpha\beta}$ describes a quadrupolar order which has been extensively studied both theoretically and experimentally in frustrated systems with ferromagnetic (FM) and antiferromagnetic (AF) nearest-neighbor and next-nearest-neighbor couplings, respectively, in strong magnetic field $h$ (see, e.g., Ref.~\cite{syromyat} and references therein) and in magnets with large non-Heisenberg spin couplings such as biquadratic exchange $({\bf S}_1{\bf S}_2)^2$. \cite{kolezh1,*kolezh2} It has been also shown recently that quantum fluctuations accompanied by a sizable single-ion easy-axis anisotropy can also stabilize a nematic phase in the kagome spin-1 antiferromagnet at $h=0$. \cite{senthil}

It is well established that the attraction between magnons caused by frustration is the origin of quadrupolar and multipolar phases in quantum magnets. \cite{chub} In particular, the bottom of the one-magnon band lies above the lowest multi-magnon bound state at $h=h_s$, where $h_s$ is the saturation field, as a result of this attraction in magnets with FM  and AF couplings between nearest- and next-nearest neighbors, respectively. Then, transitions to nematic phases at $h<h_s$ in such systems are characterized by a softening of the multi-magnon bound-state spectrum rather than the one-magnon spectrum. 

We show in the present paper that the magnon attraction arises also in spin-1 1D, 2D and 3D {\it non-frustrated} AFs with easy-axis single-ion anisotropy that leads to stabilization of nematic phases in strong magnetic field. We support our analytical results by numerical ones in the particular case of AF chain.




{\bf 2. Model and technique}. We discuss axially symmetric spin-1 systems described by the Hamiltonian
\beq
\label{ham}
\cH = \sum_{\lan i,j \ran} J_{ij} \bS_i \bS_j + D \sum_i (S^z_i)^2 - h \sum S^z_i,
\eeq
where $\langle i,j\rangle$ denote spin pairs coupled with exchange constants $J_{ij}0$, and $D<0$ is the value of the single-ion easy-axis anisotropy. 


We examine in the present paper the possibility of the nematic phase formation below the saturation field $h_s$ by considering the transition from the fully polarized state that is discussed using the Holstein-Primakoff transformation
\beq
\label{HP}
S^z_i &=& S - \akr_i a_i, \\
S^-_i &=& \akr_i \sqrt{ 2S - \akr_i a_i}. \non
\eeq
Expanding the square root in Eq.~\eqref{HP}, putting all operators $\akr_i$ to the left of all $a_i$ using commutation relations, discarding terms containing more than three operators $\akr_i$ and $a_i$ (this is reasonable because one can neglect interaction of more than two particles in a dilute gas of magnons which arises at $h\approx h_s$), \cite{bat85} and substituting the resulting expressions for $S^-_i$ and $S^+_i$ into Hamiltonian \eqref{ham} one obtains 
\beq
\cH &=& \cH_0 + \cH_2 + \cH_4 + ... \\
\label{ham2}
\cH_2 &=& \sum_\pp \left[ S J_\pp - S J_{\bf 0} - D(2S-1) + h \right] \akr_\pp a_\pp,\\
\label{ham4}
\cH_4 &=& \frac{1}{N}\sum_{1,2,3,4} \left[ D + \frac12 J_{1+3} - \cF S (J_1 + J_3) \right] \akr_1 \akr_2 a_{-3} a_{-4},\nonumber\\
\eeq
where $N$ is the number of spins, $\cF = 1 - \sqrt{1-1/2S}$, the momentum conservation law $\sum_i{\bf p}_i={\bf 0}$ is implied in Eq.~\eqref{ham4}, we omit some indexes ${\bf p}$, and we set ${\cal H}_0=0$ in the subsequent discussion.

\begin{figure}
\includegraphics[scale=0.5]{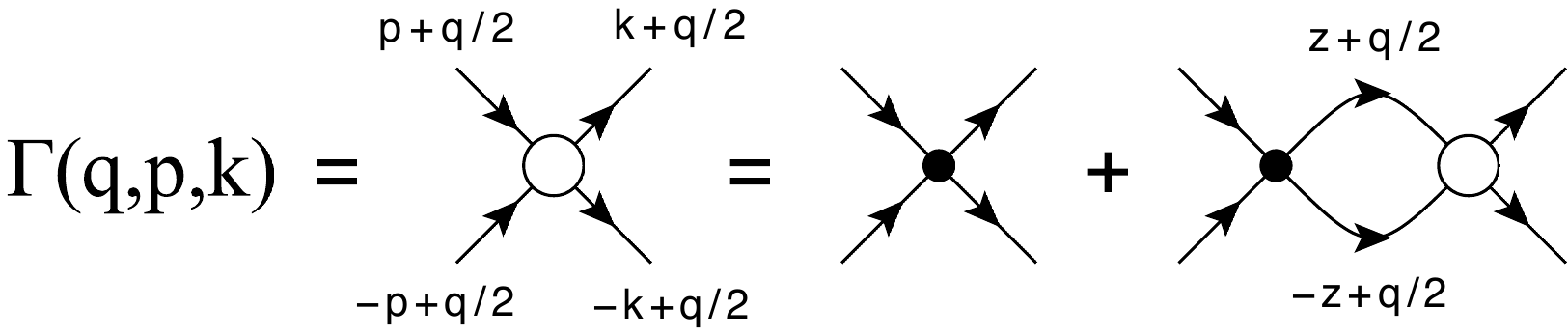}
\caption{\label{ladder} Bethe-Salpeter equation for the four-particle vertex. Black points stand for bare vertices given by Eq.~\eqref{ham4}.}
\end{figure}

At $h>h_s$, the one-magnon spectrum and the magnon Green's function are given solely by $\cH_2$, they are exact and have the form
\beq
\e_\pp &=& S J_\pp - S J_{\bf 0} - D(2S-1) + h,\\
G(\w,\pp) &=& \frac{1}{\w - \e_\pp +i\delta}.
\eeq
Two-magnon bound states are examined via analysis of the pole structure of the two-particle vertex function $\Gamma(\w, {\bf q}, {\bf p}, {\bf k})$ (see Fig.~\ref{ladder}), which can be found analytically at $h>h_s$. It can be shown that $\Gamma(\w, {\bf q}, {\bf p}, {\bf k})$ is given by a series of ladder diagrams in the fully polarized phase which lead to the Bethe-Salpeter equation for the vertex shown in Fig.~\ref{ladder}. One can represent $J_\pp$ in the following way on any Bravais lattice:
\beq
\label{jp}
J_\pp = 2J^\al \cos p_\al,
\eeq
where $\al$ enumerates exchange constants and summation over repeated Greek indexes is implied here and below. Using Eq.~\eqref{jp} we can represent the equation shown in Fig.~\ref{ladder} as follows:
\begin{eqnarray}
\label{le}
&&\Ga (\w, {\bf q}, {\bf p}, {\bf k}) = \Ga^0 (\w, {\bf q}, {\bf p}, {\bf k}) \non\\
&&-{}\int \frac{ D + J^\al \cos p_\al \cos z_\al  - 2 \cF S J^\al \cos \frac{q^\al}{2} (\cos p_\al + \cos z_\al) }
{{\cal D}(q,{\bf z})}\non\\
&&{}\times \Ga(\w, {\bf q}, {\bf z}, {\bf k})\frac{d^d{\bf z}}{(2\pi)^d},\\
&&{\cal D}(q,{\bf z}) = 2S J^\be \cos \frac{q^\be}{2} \cos z_\be - SJ_{\bf 0} - D(2S-1) + h - \frac{\w}{2}, \label{denom}\non\\
\end{eqnarray}
where $d$ is a lattice dimensionality. Let us consider the following two-particle Green's function:
\beq
\label{g2}
&&G^{II}(\w,{\bf q},\pp,{\bf k}) = \int_0^\infty dt \\
&&{}\times\langle 0| a_{\qq/2+\pp} a_{\qq/2-\pp} e^{-i (\cH-\w) t} \akr_{\qq/2+\kk} \akr_{\qq/2-\kk} |0\rangle, \non
\eeq
where $|0\rangle$ denotes the fully polarized state. Eq.~\eqref{g2} coincides with the conventional definition of the two-particle Green's function \cite{agd} 
$\langle T(a_{\qq/2+\pp}(t_1)a_{\qq/2-\pp}(t_2) \akr_{\qq/2+\kk}(t_3) \akr_{\qq/2-\kk}(t_4) )\rangle$
at $t_1=t_2=t$ and $t_3=t_4=0$, where $a_\pp(t)$ denote operators in the Heisenberg representation. If there is a two-magnon bound state $|2;\qq\rangle$ with momentum $\qq$ and energy $\e_2(\qq)$, the function $G^{II}$ has a pole at $\w=\e_2(\qq)$ near which it has the form
\beq
G^{II}(\w\sim\e_2(\qq),\qq,\pp,\kk) \approx \frac{\Phi_q(\pp)\Phi^*_q(\kk)}{\w-\e_2(\qq) + i \delta},
\eeq
where
$$\Phi_q(\pp) = \langle 2;\qq |\, \akr_{\qq/2+\pp} \akr_{\qq/2-\pp} \, | 0 \rangle$$
is a bound state wave function in momentum notation. Two-particle Green's function \eqref{g2} is equal to the vertex $\Ga(\w,{\bf q},\pp,{\bf k})$ multiplied by 

\beq  
\int d\w_p d\w_k G(p+q/2)G(-p+q/2) G(k+q/2)G(-k+q/2) \non \\
=1/(\w- \e_{\pp+{\bf q}/2} - \e_{-\pp+{\bf q}/2})(\w- \e_{\kk+{\bf q}/2} - \e_{-\kk+{\bf q}/2}). \non
\eeq 

Consequently, $\Ga(\w,{\bf q},\pp,{\bf k})$ has the form 
\beq
\label{pole}
\Ga(\w \sim \e_2({\bf q}), {\bf q}, {\bf p}, {\bf k}) \approx	\frac{ f_q(\pp)f^*_q({\bf k})}{ \w - \e_2({\bf q})+i\delta}
\eeq
and the wave function (not normalized) is related to $f_q(\pp)$ as follows:
\beq
\label{wf}
\Phi_q(\pp) = \frac{f_q(\pp)}{\e_2({\bf q}) - \e_{\pp+{\bf q}/2} - \e_{-\pp+{\bf q}/2}}.
\eeq

Certainly, it can be obtained also from Eq.~\eqref{le} that the vertex $\Ga(\w \sim \e_2({\bf q}), {\bf q}, {\bf p}, {\bf k})$ near the pole can be represented in the form \eqref{pole}. It is also evident from Eq.~\eqref{le} that $f_q(\pp)$ has the form
\beq
\label{fdecomp}
f_q(\pp) = A_q + B_q^\al \cos p_\al.
\eeq
Substituting Eqs.~\eqref{pole} and \eqref{fdecomp} into Eq.~\eqref{le} and neglecting $\Ga^0 (\w, {\bf q}, {\bf p}, {\bf k})$ we obtain the following set of equations for $A_q$ and $B_q$:
\beq
\label{matrix_eq}
M(q,D,h,\w) \left( \begin{array}{l} A_q \\ B_q^{\al=1} \\ B_q^{\al=2} \\...\end{array} \right) = 0,
\eeq
where $M(q,D,h,\w)$ is a square matrix with the following elements:
\beq
M_{00} &=& 1 + D g - 2 \cF S J^\al \cos \frac{q^\al}{2} v_\al, \non\\
M_{0\al} &=& D v_\al - 2 \cF S J^\be \cos \frac{q^\be}{2} I_{\be\al}, \non\\
M_{\al0} &=& J_\al v_\al - 2 g \cF S J_\al \cos \frac{q_\al}{2}, \\
M_{\al\be} &=& \delta_{\al\be} + J_\al I_{\al\be} - 2\cF SJ_\al \cos \frac{q_\al}{2} v_\be, \non
\eeq
where
\beq
g &=& \int\frac{d^d{\bf z}}{(2\pi)^d}\frac{1}{{\cal D}(q,{\bf z})}, \non\\
v^\al &=& \int\frac{d^d{\bf z}}{(2\pi)^d} \frac{\cos z^\al}{{\cal D}(q,{\bf z})},\label{ints}\\ 
I^{\al\be} &=& \int\frac{d^d{\bf z}}{(2\pi)^d} \frac{\cos z^\al \cos z^\be}{{\cal D}(q,{\bf z})},\non
\eeq
and ${\cal D}(q,{\bf z})$ is given by Eq.~\eqref{denom}. Then, if determinant of $M$ is zero at some $\w$, Eq.~\eqref{matrix_eq} has a non-trivial solution and the vertex has a pole at the corresponding $\w$ and $\bf q$. Otherwise, the residue is zero and there is no bound state at the corresponding $\w$ and $\bf q$. 

One notes that parameters \eqref{ints} depend on $D,\,h$ and $\w$ only in the combination
$$ \tilh = h - \w/2 - D(2S-1)$$
and $M(q,D,h,\w)$ is actually a function of $(D,\tilh)$. Considering $D$ and $\tilh$ as independent variables, we note that $\det M(D,\tilh)$ is a {\it linear} function of $D$. It signifies that there exists a unique $D$ value at which $\det M(D,\tilh)=0$ for any $\tilh$ larger than some value $\tilh_0$ such that denominator \eqref{denom} is positive at $\tilh>\tilh_0$. Then, the two-magnon bound state exists at $\tilh>\tilh_0$ and the critical field at which the two-magnon bound state spectrum becomes gapless is {\it higher} than the critical field of the one-magnon mode (if these critical fields exist \footnote{In a range of parameters bound states are gapped at any $h$}). 

The above equations are applicable for consideration of two-magnon bound states in strong magnetic field on any Bravais lattice with any exchange interactions and any $S \geqslant1$. They can be easily extended also to the case of anisotropic exchange. We apply the above equations now to the particular spin-1 model on a tetragonal lattice with different unfrustrated exchange constants along $z$-axis ($J^z$) and in $xy$ plane ($J^{xy}$) which are parametrized as follows: 
\beq
\label{param}
J^z &=& \cos \theta, \non\\
J^{xy} &=& \sin \theta,  \\
-\pi < & \theta & \leqslant \pi. \non
\eeq
The cases of $\theta = 0$ and $\pi$ correspond to 1D AF and FM chains, respectively, $\theta = \pm \pi/2$ correspond to 2D AFs and FMs on the square lattice. Other $\theta$ describe axially symmetric 3D magnets.

\begin{figure}
\includegraphics[scale=0.4]{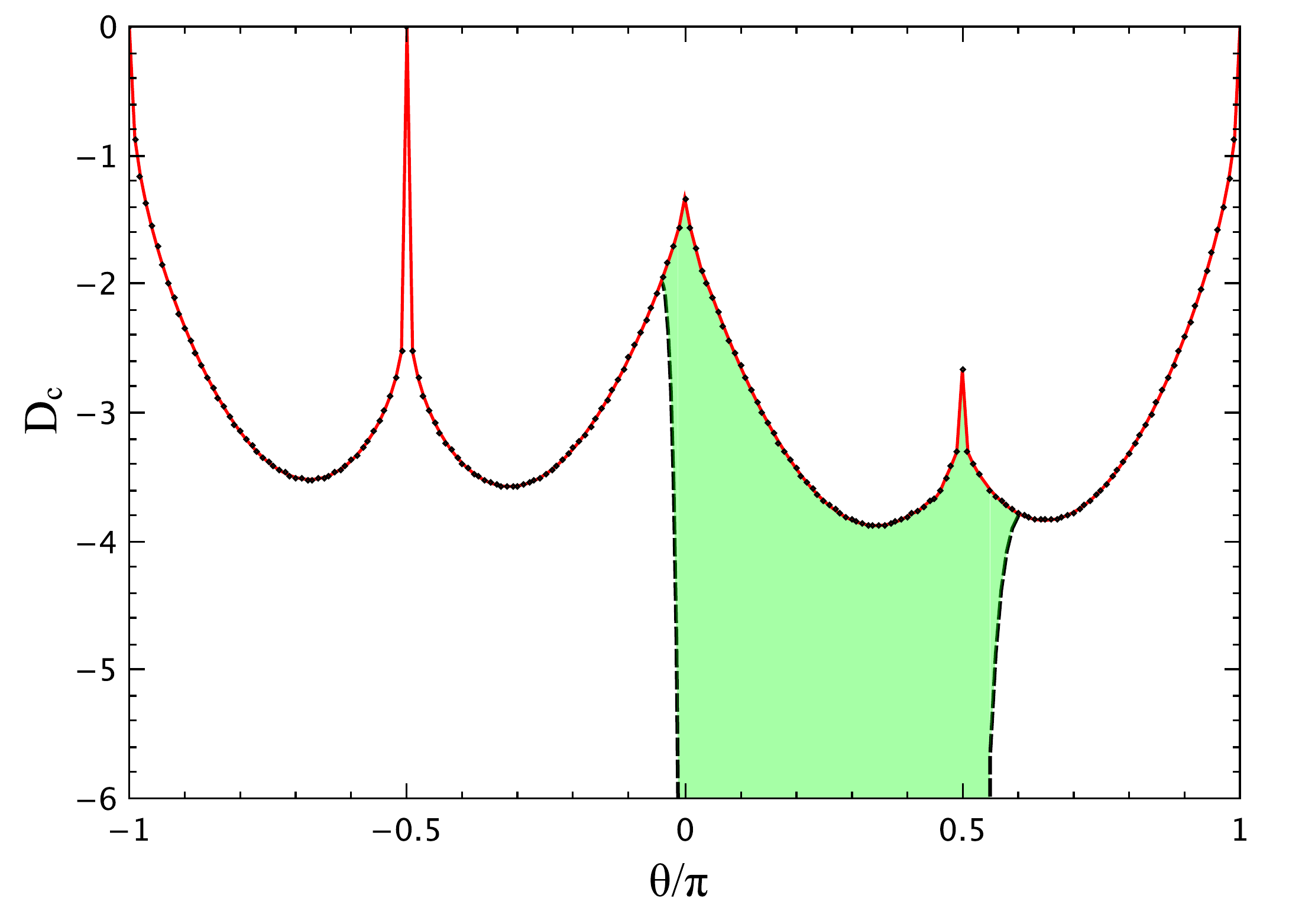}
\caption{\label{dc_graph} Solid red line (black dots) represents $D_c (\theta)$ curve (see text). Colored area is the nematic phase stability region. }
\end{figure}

{\bf 3. General consideration.} One has $J_\pp = 2J_z \cos p_z + 2J_{xy} (\cos p_x + \cos p_y)$ and we find using the above formulas that the bound state at strong field exists at any $\theta$ if $D<D_c(\theta)$. The bound state spectrum is always quadratic near it's minimum $\pp={\bf 0}$: $\e_2(\pp) \approx 2 (h-h_s) + C_z p_z^2 + C_{xy} (p_x^2 + p_y^2)$. Function $D_c(\theta)$ is shown in Fig.~\ref{dc_graph} by red solid line. Behavior of $D_c(\theta)$ near its singularities which correspond to 1D and 2D systems can be found using \eqref{ints} with the following result:
\beq
D_c(\theta \sim 0) &\approx& -4/3 + {\rm const} \times \sqrt{|\theta|}, \non \\
D_c(\theta \sim \pi/2) &\approx& -8/3 + \frac{{\rm const}}{\ln |\theta - \pi/2|}, \non \\
D_c(\theta \sim -\pi/2) &\approx& \frac{{\rm const}}{\ln |\theta - \pi/2|}, \non \\
D_c(\theta \sim \pm\pi) &\approx& {\rm const} \times \sqrt{|\theta\mp\pi|}. \non 
\eeq
The bound state can formally become gapless upon the field decreasing if $\theta$ lies in the range $-0.197 \pi < \theta < 0.905 \pi$. In this interval, Bose condensate of bound state quasiparticles would appear below the critical field. But a spin-flop transition at certain values of $D$ occurs at field values which are higher than the critical field of the bound state. Detailed consideration of the spin-flop transition is out of the aim of the present paper. We restrict ourselves by a simple classical result for the spin-flop field obtained by minimization of the classical energy. According to the classical result, the spin-flop transition occurs before condensation of bound states at $D<D_{sf1}(\theta),\, \theta<0$ and $D<D_{sf2}(\theta),\, \theta>\pi/2$, where $D_{sf1}(\theta)$ and $D_{sf2}(\theta)$ are shown in Fig.~\ref{dc_graph} by dashed black lines. They approach $-\infty$ as $\theta \rightarrow -0$ and $\theta \rightarrow \pi/2+0$, correspondingly. 
As a result the area of the nematic phase stability turns out to be narrower and it is shown in Fig.~\ref{dc_graph}. 

Assuming repulsion interaction between bound states quasiparticles near its spectrum minimum (that is reasonable due to antiferromagnetic interaction) we obtain for the nematic phase at $h \lesssim h_s$ and $\theta\ne0$:
\beq
\label{sxy}
\lan S^x_i \ran = \lan S^y_i \ran &=& 0, \\
\label{cond}
\left\lan \sum_j \Phi_0(r_j-r_i) S^+_i S^+_j \right\ran & = & 2 e^{i\phi} \sqrt{N\rho},
\eeq
and
\beq
\label{sz}
1 - \lan S^z_i \ran = 2 \rho &\approx& \frac{ 4 ( h_s-h ) }{\la}, \quad \theta \ne \pi/2 \non \\
h_s - h &\approx& -\frac{4 \pi C_{xy} \rho}{\ln \rho},\quad  \theta = \pi/2. \non
\eeq 
where $\Phi_0(r)$ is the wave function of the bound state \eqref{wf} with zero momentum in the spatial representation, $\rho$ is the condensate density, and $\la$ is an effective coupling of bound state quasiparticles near the spectrum minimum (in 3D case). Thus, we obtain a long-range nematic order (Eq.~\eqref{cond}) without a conventional magnetic order in $xy$ plane (Eq.~\eqref{sxy}). Due to the quadratic dispersion of $\e_2(\pp)$ we also have the 3D BEC relation \cite{syromyat}
\beq
h_s(0) - h_s(T) \propto T^{3/2}.
\eeq

\begin{figure}
\includegraphics[scale=0.43]{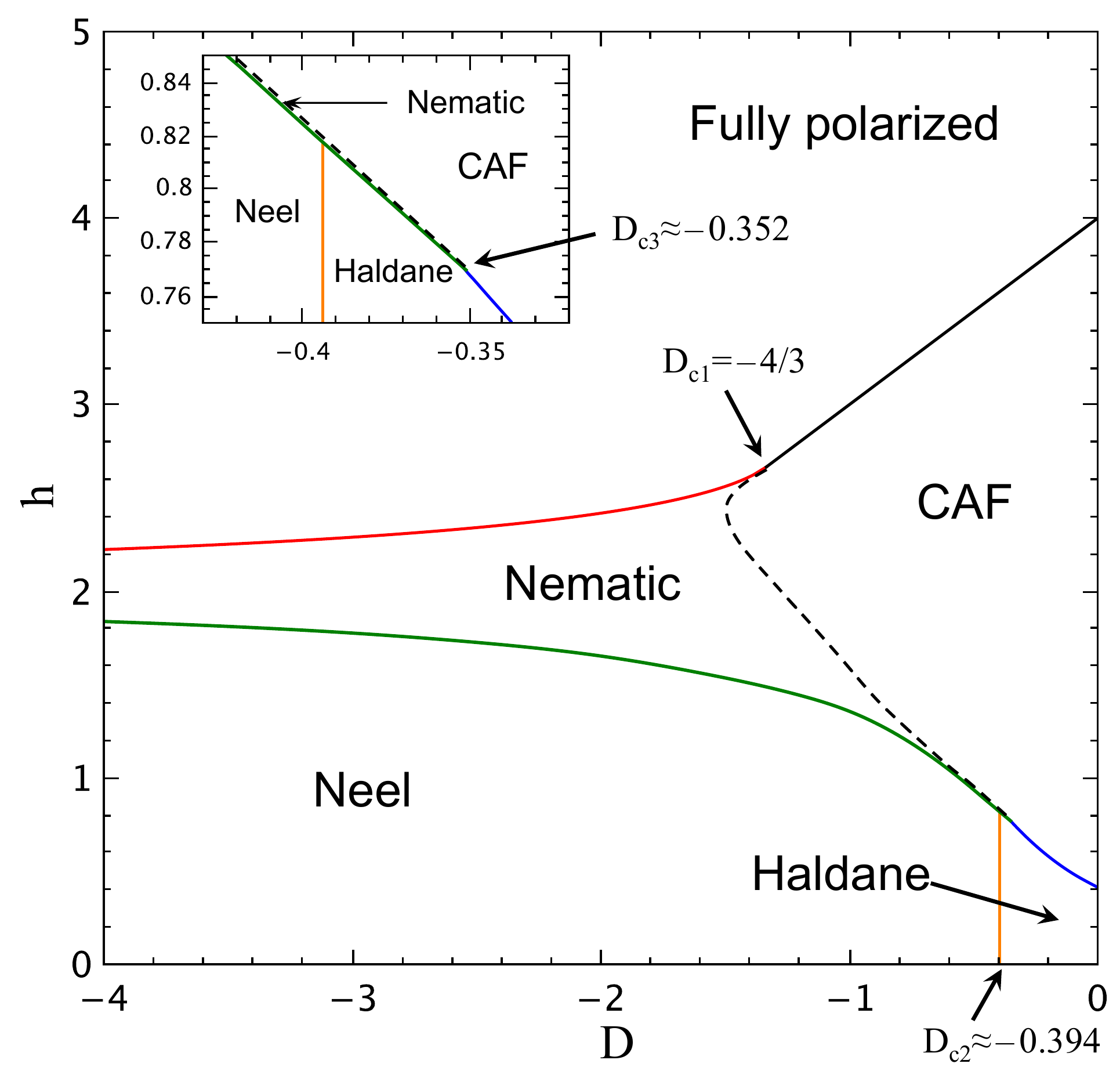}
\caption{\label{phase} Phase diagram of the isolated chain. The Neel and Haldane phases are gapped, while the nematic and canted antiferromagnetic (CAF) ones are gapless. 
Inset shows the region, where four phases meet. }
\end{figure}

{\bf 4. 1D AF chain ($\theta = 0$).} 
One has in this case $J_\pp = 2 \cos p$ and all results (saturation field, $C_z$, etc.) can be found analytically which however turn out to be quite cumbersome. Then, we do not present all of them here. As it is mentioned above, bound state exist at $D<D_{c1}=-4/3$. The saturation field $h_s$ is shown in Fig.~\ref{phase} as a function of $D$. 
Bearing in mind the quadratic dispersion of $\e_2(p)$ near its minimum and using results of 1D Bose-gas discussions \cite{lieb,*lieb2,*kor} one obtains \cite{see} at $h \lesssim h_s$ and $T=0$
\begin{eqnarray}
\label{mag1d}
	&& 1 -  \left\lan S_j^z \right\ran   = 2 \rho = \frac2\pi \sqrt{\frac{2(h_s-h)}{C}},\\
\label{zz}
&&\left\langle S_{j+n}^z(t) S_{j}^z(0) \right\rangle 
\approx 
\left\langle S_{j}^z \right\rangle^2 
\\
&&- \frac{1}{\pi}\left(\frac{1}{(n+iut)^2} + \frac{1}{(n-iut)^2}\right) 
+ B_1\frac{\cos(\pi\langle a_j^\dagger a_j \rangle n)}{n^2+u^2t^2},\nonumber\\
\label{nemcorr}
&&\left\langle \left(S_{0}^+(t)\right)^2 \left(S_{n}^-(0)\right)^2  \right\rangle 
\approx \frac{B_2}{\sqrt{|n+iut|}},
\end{eqnarray}
where $\rho$ is the density of quasiparticles describing bound states, $n\gg1$, $u=4\pi C_z \rho$ and $B_{1,2}$ are constants. All other correlators are either exponentially small or equal to zero exactly. Then, the nematic phase in the isolated chain is characterized by no long-range magnetic order and quasi-long-range nematic order described by the algebraic decay of the nematic correlator \eqref{nemcorr}.

To support our analytical results we have performed numerical analysis using exact diagonalization application from ALPS software package. \cite{alps2} This numerical consideration is simplified by the fact that the Hamiltonian \eqref{ham} commutes with the $z$ component of the total spin ${\cal S}^z$ and with the Zeeman term. As a result all the Hamiltonian eigenstates can be classified by eigenvalues $M$ of ${\cal S}^z$. Let us denote $E(M)$ the minimum energy in each $M$ sector at $h=0$. The ground state energy of a cluster with $L$ spins in magnetic field is given by the minimum value of $E(M)-hM$. An important observation is that values $E(M)-h M$ at even $M_{sat}-M$ are smaller than those with odd $M_{sat}-M$, where $M_{sat}=L$. This property holds for different sizes ($L = 6 \div 16 $) and both periodic (for even $L$) and open boundary conditions.

Thus, one can expect that a quasi-condensation takes place in the thermodynamical limit of elementary excitations carrying spin 2. Values of $h_s$ obtained numerically as a result of analysis of clusters with $L=8\div40$ are in excellent agreement with the corresponding values found analytically. Numerical consideration of clusters with $L=8\div24$ similar to that performed in Refs.~\cite{sakai1,sakai2} confirms also the validity of Eq.~\eqref{mag1d}. For example, one obtains for $D=-3.5$ that $1 - \left\langle S_j^z\right\rangle = \alpha(h_s-h)^{1/\delta}$, where $\alpha=6.18\pm 0.15$ and $\delta=2.0041\pm0.003$, that is in good agreement with the analytical result $1 - \left\langle S_j^z\right\rangle \approx 6.37\sqrt{H_s-H}$. We have calculated numerically magnetization near saturation for several $D$ values and found results which differ from the theory by a few percents only. 
 
The phase diagram obtained numerically is shown in Fig.~\ref{phase} that is in very good agreement with previous numerical discussions of the present 1D system \cite{sakai0,1dd5,jap1}. The line of phase transitions from the fully polarized state is obtained by consideration of some small-$(M_{sat}-M)$ sectors of clusters with $L = 8 \div 40$. This transition takes place to the nematic and CAF phases at $D<D_{c1}=-4/3$ and $D>D_{c1}$, respectively, that is in agreement with our theoretical predictions. 

At small $h$ the system goes from the Haldane phase to that with Neel long-range order as the easy-axis anisotropy rises. The transition point $D=D_{c2}\approx-0.394$ is found using the extrapolating procedure from $L=8,10,12,14,16$ as it is explained in section III.B of Ref.~\cite{full_phase}. The value $D_{c2}$ does not depend on $h$ because both Neel and Haldane phases have gaps and their ground state energy do not depend on $h$.

Phase transitions take place upon the field increasing from Neel and Haldane phases to nematic and CAF ones. We observe that the lowest excited states belong either to $M=1$ or $M=2$ sectors in Haldane and Neel phases. Then, corresponding critical fields $h_c(M)$ can be found using the value $h_c(M,L) = \frac{E(M,L)-E(0,L)}{M}$ as $h_c(M,L\to\infty)=h_c(M)$. If $h_c(1)<(>)h_c(2)$ the transition takes place to CAF (nematic) phase. For each $D$ value, $h_c(M,L)$ have been found for $L=8$, 10, 12, 14 and $h_c(1)$, $h_c(2)$ have been obtained by an extrapolation \cite{extrapol} of $h_c(M,L)$ to the thermodynamical limit $L\to\infty$. We have obtained that $h_c(1)$ and $h_c(2)$ intersect at $D_{c3} \approx -0.352$ (see Fig.~\ref{phase}). Two multi-critical points are shown in Fig.~\ref{phase} which are close to each other: $D_{c2} \approx -0.394 \pm 0.01 $, $h_{c2} \approx 0.817 \pm 0.013 $ and $D_{c3} \approx -0.352\pm0.003$, $h_{c3} \approx 0.772\pm0.004$. However, the precision of our calculations is insufficient to exclude the possibility of these two points merging into a single multi-critical one. The boundary between the nematic and CAF phases in Fig.~\ref{phase} (dashed line) is taken from Ref.~\cite{sakai0} 

{\bf 5. Summary and conclusion.} We demonstrate in the present paper that easy-axis single-ion anisotropy $D$ in non-frustrated spin-1 antiferromagnetic systems can lead to spin-nematic states in strong magnetic field $h$. General formulas are derived for consideration of transitions in such systems from the fully polarized phase to the nematic one. The region of the nematic phase stability is obtained in the model \eqref{ham} with exchange constants \eqref{param} parametrized by the value $\theta$ (see Fig.~\ref{dc_graph}). We discuss in detail both analytically and numerically the special case of $\theta=0$ that describes AF chain. We obtain the phase diagram of the isolated chain on $h-D$ plane (see Fig.~\ref{phase}). We hope that the present study will stimulate further theoretical and experimental activity in nematic phases discussion both in the considered systems and in other ones containing anisotropy.

This work was supported by the President of the Russian Federation (Grant No.\ MD-274.2012.2), the Dynasty foundation and RFBR Grants No.\ 12-02-01234 and No.\ 12-02-00498.

\bibliography{nemlib} 

\begin{thebibliography}{21}%
\makeatletter
\providecommand \@ifxundefined [1]{%
 \@ifx{#1\undefined}
}%
\providecommand \@ifnum [1]{%
 \ifnum #1\expandafter \@firstoftwo
 \else \expandafter \@secondoftwo
 \fi
}%
\providecommand \@ifx [1]{%
 \ifx #1\expandafter \@firstoftwo
 \else \expandafter \@secondoftwo
 \fi
}%
\providecommand \natexlab [1]{#1}%
\providecommand \enquote  [1]{``#1''}%
\providecommand \bibnamefont  [1]{#1}%
\providecommand \bibfnamefont [1]{#1}%
\providecommand \citenamefont [1]{#1}%
\providecommand \href@noop [0]{\@secondoftwo}%
\providecommand \href [0]{\begingroup \@sanitize@url \@href}%
\providecommand \@href[1]{\@@startlink{#1}\@@href}%
\providecommand \@@href[1]{\endgroup#1\@@endlink}%
\providecommand \@sanitize@url [0]{\catcode `\\12\catcode `\$12\catcode
  `\&12\catcode `\#12\catcode `\^12\catcode `\_12\catcode `\%12\relax}%
\providecommand \@@startlink[1]{}%
\providecommand \@@endlink[0]{}%
\providecommand \url  [0]{\begingroup\@sanitize@url \@url }%
\providecommand \@url [1]{\endgroup\@href {#1}{\urlprefix }}%
\providecommand \urlprefix  [0]{URL }%
\providecommand \Eprint [0]{\href }%
\providecommand \doibase [0]{http://dx.doi.org/}%
\providecommand \selectlanguage [0]{\@gobble}%
\providecommand \bibinfo  [0]{\@secondoftwo}%
\providecommand \bibfield  [0]{\@secondoftwo}%
\providecommand \translation [1]{[#1]}%
\providecommand \BibitemOpen [0]{}%
\providecommand \bibitemStop [0]{}%
\providecommand \bibitemNoStop [0]{.\EOS\space}%
\providecommand \EOS [0]{\spacefactor3000\relax}%
\providecommand \BibitemShut  [1]{\csname bibitem#1\endcsname}%
\let\auto@bib@innerbib\@empty
\bibitem [{\citenamefont {Andreev}\ and\ \citenamefont
  {Grishchuk}(1984)}]{andreev}%
  \BibitemOpen
  \bibfield  {author} {\bibinfo {author} {\bibfnamefont {A.~F.}\ \bibnamefont
  {Andreev}}\ and\ \bibinfo {author} {\bibfnamefont {I.~A.}\ \bibnamefont
  {Grishchuk}},\ }\href@noop {} {\bibfield  {journal} {\bibinfo  {journal}
  {Sov. Phys. JETP}\ }\textbf {\bibinfo {volume} {60}},\ \bibinfo {pages} {267}
  (\bibinfo {year} {1984})}\BibitemShut {NoStop}%
\bibitem [{\citenamefont {Syromyatnikov}(2012)}]{syromyat}%
  \BibitemOpen
  \bibfield  {author} {\bibinfo {author} {\bibfnamefont {A.~V.}\ \bibnamefont
  {Syromyatnikov}},\ }\href@noop {} {\bibfield  {journal} {\bibinfo  {journal}
  {Phys. Rev. B}\ }\textbf {\bibinfo {volume} {86}},\ \bibinfo {pages} {014423}
  (\bibinfo {year} {2012})}\BibitemShut {NoStop}%
\bibitem [{\citenamefont {Rodriguez}\ \emph {et~al.}(2011)\citenamefont
  {Rodriguez}, \citenamefont {Arg\"uelles}, \citenamefont {Kolezhuk},
  \citenamefont {Santos},\ and\ \citenamefont {Vekua}}]{kolezh1}%
  \BibitemOpen
  \bibfield  {author} {\bibinfo {author} {\bibfnamefont {K.}~\bibnamefont
  {Rodriguez}}, \bibinfo {author} {\bibfnamefont {A.}~\bibnamefont
  {Arg\"uelles}}, \bibinfo {author} {\bibfnamefont {A.~K.}\ \bibnamefont
  {Kolezhuk}}, \bibinfo {author} {\bibfnamefont {L.}~\bibnamefont {Santos}}, \
  and\ \bibinfo {author} {\bibfnamefont {T.}~\bibnamefont {Vekua}},\
  }\href@noop {} {\bibfield  {journal} {\bibinfo  {journal} {Phys. Rev. Lett.}\
  }\textbf {\bibinfo {volume} {106}},\ \bibinfo {pages} {105302} (\bibinfo
  {year} {2011})}\BibitemShut {NoStop}%
\bibitem [{\citenamefont {Fridman}\ \emph {et~al.}(2011)\citenamefont
  {Fridman}, \citenamefont {Kosmachev}, \citenamefont {Kolezhuk},\ and\
  \citenamefont {Ivanov}}]{kolezh2}%
  \BibitemOpen
  \bibfield  {author} {\bibinfo {author} {\bibfnamefont {Y.~A.}\ \bibnamefont
  {Fridman}}, \bibinfo {author} {\bibfnamefont {O.~A.}\ \bibnamefont
  {Kosmachev}}, \bibinfo {author} {\bibfnamefont {A.~K.}\ \bibnamefont
  {Kolezhuk}}, \ and\ \bibinfo {author} {\bibfnamefont {B.~A.}\ \bibnamefont
  {Ivanov}},\ }\href@noop {} {\bibfield  {journal} {\bibinfo  {journal} {Phys.
  Rev. Lett.}\ }\textbf {\bibinfo {volume} {106}},\ \bibinfo {pages} {097202}
  (\bibinfo {year} {2011})}\BibitemShut {NoStop}%
\bibitem [{\citenamefont {Damle}\ and\ \citenamefont
  {Senthil}(2006)}]{senthil}%
  \BibitemOpen
  \bibfield  {author} {\bibinfo {author} {\bibfnamefont {K.}~\bibnamefont
  {Damle}}\ and\ \bibinfo {author} {\bibfnamefont {T.}~\bibnamefont
  {Senthil}},\ }\href@noop {} {\bibfield  {journal} {\bibinfo  {journal} {Phys.
  Rev. Lett.}\ }\textbf {\bibinfo {volume} {97}},\ \bibinfo {pages} {067202}
  (\bibinfo {year} {2006})}\BibitemShut {NoStop}%
\bibitem [{\citenamefont {Chubukov}(1991)}]{chub}%
  \BibitemOpen
  \bibfield  {author} {\bibinfo {author} {\bibfnamefont {A.~V.}\ \bibnamefont
  {Chubukov}},\ }\href {\doibase 10.1103/PhysRevB.44.4693} {\bibfield
  {journal} {\bibinfo  {journal} {Phys. Rev. B}\ }\textbf {\bibinfo {volume}
  {44}},\ \bibinfo {pages} {4693} (\bibinfo {year} {1991})}\BibitemShut
  {NoStop}%
\bibitem [{\citenamefont {Batyev}(1985)}]{bat85}%
  \BibitemOpen
  \bibfield  {author} {\bibinfo {author} {\bibfnamefont {E.~G.}\ \bibnamefont
  {Batyev}},\ }\href@noop {} {\bibfield  {journal} {\bibinfo  {journal} {Sov.
  Phys. JETP}\ }\textbf {\bibinfo {volume} {62}},\ \bibinfo {pages} {173}
  (\bibinfo {year} {1985})}\BibitemShut {NoStop}%
\bibitem [{\citenamefont {Abrikosov}\ \emph {et~al.}(1963)\citenamefont
  {Abrikosov}, \citenamefont {Gor'kov},\ and\ \citenamefont
  {Dzyaloshinskii}}]{agd}%
  \BibitemOpen
  \bibfield  {author} {\bibinfo {author} {\bibfnamefont {A.~A.}\ \bibnamefont
  {Abrikosov}}, \bibinfo {author} {\bibfnamefont {L.~P.}\ \bibnamefont
  {Gor'kov}}, \ and\ \bibinfo {author} {\bibfnamefont {I.~E.}\ \bibnamefont
  {Dzyaloshinskii}},\ }\href@noop {} {\emph {\bibinfo {title} {Quantum Field
  Theoretical Methods in Statistical Physics}}}\ (\bibinfo  {publisher}
  {Dover},\ \bibinfo {address} {New York},\ \bibinfo {year} {1963})\BibitemShut
  {NoStop}%
\bibitem [{Note1()}]{Note1}%
  \BibitemOpen
  \bibinfo {note} {In a range of parameters bound states are gapped at any
  $h$}\BibitemShut {NoStop}%
\bibitem [{\citenamefont {Lieb}\ and\ \citenamefont {Liniger}(1963)}]{lieb}%
  \BibitemOpen
  \bibfield  {author} {\bibinfo {author} {\bibfnamefont {E.~H.}\ \bibnamefont
  {Lieb}}\ and\ \bibinfo {author} {\bibfnamefont {W.}~\bibnamefont {Liniger}},\
  }\href {\doibase 10.1103/PhysRev.130.1605} {\bibfield  {journal} {\bibinfo
  {journal} {Phys. Rev.}\ }\textbf {\bibinfo {volume} {130}},\ \bibinfo {pages}
  {1605} (\bibinfo {year} {1963})}\BibitemShut {NoStop}%
\bibitem [{\citenamefont {Lieb}(1963)}]{lieb2}%
  \BibitemOpen
  \bibfield  {author} {\bibinfo {author} {\bibfnamefont {E.~H.}\ \bibnamefont
  {Lieb}},\ }\href {\doibase 10.1103/PhysRev.130.1616} {\bibfield  {journal}
  {\bibinfo  {journal} {Phys. Rev.}\ }\textbf {\bibinfo {volume} {130}},\
  \bibinfo {pages} {1616} (\bibinfo {year} {1963})}\BibitemShut {NoStop}%
\bibitem [{\citenamefont {Korepin}\ \emph {et~al.}(1993)\citenamefont
  {Korepin}, \citenamefont {Bogoliubov},\ and\ \citenamefont {Izergin}}]{kor}%
  \BibitemOpen
  \bibfield  {author} {\bibinfo {author} {\bibfnamefont {V.~E.}\ \bibnamefont
  {Korepin}}, \bibinfo {author} {\bibfnamefont {N.~M.}\ \bibnamefont
  {Bogoliubov}}, \ and\ \bibinfo {author} {\bibfnamefont {A.~G.}\ \bibnamefont
  {Izergin}},\ }\href@noop {} {\emph {\bibinfo {title} {Quantum Inverse
  Scattering Method and Correlation Functions}}}\ (\bibinfo  {publisher}
  {Cambridge University Press},\ \bibinfo {address} {Cambridge},\ \bibinfo
  {year} {1993})\BibitemShut {NoStop}%
\bibitem [{see()}]{see}%
  \BibitemOpen
  \href@noop {} {}\bibinfo {note} {See Ref.~\cite{syromyat} for an extended
  discussion}\BibitemShut {NoStop}%
\bibitem [{\citenamefont {Bauer}\ \emph {et~al.}(2011)\citenamefont {Bauer},
  \citenamefont {Carr}, \citenamefont {Evertz}, \citenamefont {Feiguin},
  \citenamefont {Freire}, \citenamefont {Fuchs}, \citenamefont {Gamper},
  \citenamefont {Gukelberger}, \citenamefont {Gull}, \citenamefont {Guertler},
  \citenamefont {Hehn}, \citenamefont {Igarashi}, \citenamefont {Isakov},
  \citenamefont {Koop}, \citenamefont {Ma}, \citenamefont {Mates},
  \citenamefont {Matsuo}, \citenamefont {Parcollet}, \citenamefont {Paw?owski},
  \citenamefont {Picon}, \citenamefont {Pollet}, \citenamefont {Santos},
  \citenamefont {Scarola}, \citenamefont {Schollw?ck}, \citenamefont {Silva},
  \citenamefont {Surer}, \citenamefont {Todo}, \citenamefont {Trebst},
  \citenamefont {Troyer}, \citenamefont {Wall}, \citenamefont {Werner},\ and\
  \citenamefont {Wessel}}]{alps2}%
  \BibitemOpen
  \bibfield  {author} {\bibinfo {author} {\bibfnamefont {B.}~\bibnamefont
  {Bauer}}, \bibinfo {author} {\bibfnamefont {L.~D.}\ \bibnamefont {Carr}},
  \bibinfo {author} {\bibfnamefont {H.~G.}\ \bibnamefont {Evertz}}, \bibinfo
  {author} {\bibfnamefont {A.}~\bibnamefont {Feiguin}}, \bibinfo {author}
  {\bibfnamefont {J.}~\bibnamefont {Freire}}, \bibinfo {author} {\bibfnamefont
  {S.}~\bibnamefont {Fuchs}}, \bibinfo {author} {\bibfnamefont
  {L.}~\bibnamefont {Gamper}}, \bibinfo {author} {\bibfnamefont
  {J.}~\bibnamefont {Gukelberger}}, \bibinfo {author} {\bibfnamefont
  {E.}~\bibnamefont {Gull}}, \bibinfo {author} {\bibfnamefont {S.}~\bibnamefont
  {Guertler}}, \bibinfo {author} {\bibfnamefont {A.}~\bibnamefont {Hehn}},
  \bibinfo {author} {\bibfnamefont {R.}~\bibnamefont {Igarashi}}, \bibinfo
  {author} {\bibfnamefont {S.~V.}\ \bibnamefont {Isakov}}, \bibinfo {author}
  {\bibfnamefont {D.}~\bibnamefont {Koop}}, \bibinfo {author} {\bibfnamefont
  {P.~N.}\ \bibnamefont {Ma}}, \bibinfo {author} {\bibfnamefont
  {P.}~\bibnamefont {Mates}}, \bibinfo {author} {\bibfnamefont
  {H.}~\bibnamefont {Matsuo}}, \bibinfo {author} {\bibfnamefont
  {O.}~\bibnamefont {Parcollet}}, \bibinfo {author} {\bibfnamefont
  {G.}~\bibnamefont {Paw?owski}}, \bibinfo {author} {\bibfnamefont {J.~D.}\
  \bibnamefont {Picon}}, \bibinfo {author} {\bibfnamefont {L.}~\bibnamefont
  {Pollet}}, \bibinfo {author} {\bibfnamefont {E.}~\bibnamefont {Santos}},
  \bibinfo {author} {\bibfnamefont {V.~W.}\ \bibnamefont {Scarola}}, \bibinfo
  {author} {\bibfnamefont {U.}~\bibnamefont {Schollw?ck}}, \bibinfo {author}
  {\bibfnamefont {C.}~\bibnamefont {Silva}}, \bibinfo {author} {\bibfnamefont
  {B.}~\bibnamefont {Surer}}, \bibinfo {author} {\bibfnamefont
  {S.}~\bibnamefont {Todo}}, \bibinfo {author} {\bibfnamefont {S.}~\bibnamefont
  {Trebst}}, \bibinfo {author} {\bibfnamefont {M.}~\bibnamefont {Troyer}},
  \bibinfo {author} {\bibfnamefont {M.~L.}\ \bibnamefont {Wall}}, \bibinfo
  {author} {\bibfnamefont {P.}~\bibnamefont {Werner}}, \ and\ \bibinfo {author}
  {\bibfnamefont {S.}~\bibnamefont {Wessel}},\ }\href@noop {} {\bibfield
  {journal} {\bibinfo  {journal} {Journal of Statistical Mechanics: Theory and
  Experiment}\ }\textbf {\bibinfo {volume} {2011}},\ \bibinfo {pages} {P05001}
  (\bibinfo {year} {2011})}\BibitemShut {NoStop}%
\bibitem [{\citenamefont {Sakai}\ and\ \citenamefont
  {Takahashi}(1991)}]{sakai1}%
  \BibitemOpen
  \bibfield  {author} {\bibinfo {author} {\bibfnamefont {T.}~\bibnamefont
  {Sakai}}\ and\ \bibinfo {author} {\bibfnamefont {M.}~\bibnamefont
  {Takahashi}},\ }\href@noop {} {\bibfield  {journal} {\bibinfo  {journal}
  {Phys. Rev. B}\ }\textbf {\bibinfo {volume} {43}},\ \bibinfo {pages} {13383}
  (\bibinfo {year} {1991})}\BibitemShut {NoStop}%
\bibitem [{\citenamefont {Sakai}\ and\ \citenamefont
  {Takahashi}(1998)}]{sakai2}%
  \BibitemOpen
  \bibfield  {author} {\bibinfo {author} {\bibfnamefont {T.}~\bibnamefont
  {Sakai}}\ and\ \bibinfo {author} {\bibfnamefont {M.}~\bibnamefont
  {Takahashi}},\ }\href@noop {} {\bibfield  {journal} {\bibinfo  {journal}
  {Phys. Rev. B}\ }\textbf {\bibinfo {volume} {57}},\ \bibinfo {pages} {R8091}
  (\bibinfo {year} {1998})}\BibitemShut {NoStop}%
\bibitem [{\citenamefont {Sakai}(1998)}]{sakai0}%
  \BibitemOpen
  \bibfield  {author} {\bibinfo {author} {\bibfnamefont {T.}~\bibnamefont
  {Sakai}},\ }\href@noop {} {\bibfield  {journal} {\bibinfo  {journal} {Phys.
  Rev. B}\ }\textbf {\bibinfo {volume} {58}},\ \bibinfo {pages} {6268}
  (\bibinfo {year} {1998})}\BibitemShut {NoStop}%
\bibitem [{\citenamefont {Tonegawa}\ \emph {et~al.}(1996)\citenamefont
  {Tonegawa}, \citenamefont {Nakao},\ and\ \citenamefont {Kaburagi}}]{1dd5}%
  \BibitemOpen
  \bibfield  {author} {\bibinfo {author} {\bibfnamefont {T.}~\bibnamefont
  {Tonegawa}}, \bibinfo {author} {\bibfnamefont {T.}~\bibnamefont {Nakao}}, \
  and\ \bibinfo {author} {\bibfnamefont {M.}~\bibnamefont {Kaburagi}},\
  }\href@noop {} {\bibfield  {journal} {\bibinfo  {journal} {J. Phys. Soc.
  Jpn.}\ }\textbf {\bibinfo {volume} {65}},\ \bibinfo {pages} {3317} (\bibinfo
  {year} {1996})}\BibitemShut {NoStop}%
\bibitem [{\citenamefont {Takashi~Tonegawa}\ and\ \citenamefont
  {Kaburagi}(2005)}]{jap1}%
  \BibitemOpen
  \bibfield  {author} {\bibinfo {author} {\bibfnamefont {T.~S.}\ \bibnamefont
  {Takashi~Tonegawa}, \bibfnamefont {Kouichi~Okunishi}}\ and\ \bibinfo {author}
  {\bibfnamefont {M.}~\bibnamefont {Kaburagi}},\ }\href@noop {} {\bibfield
  {journal} {\bibinfo  {journal} {Progr. Theor. Phys. Suppl.}\ }\textbf
  {\bibinfo {volume} {159}},\ \bibinfo {pages} {77} (\bibinfo {year}
  {2005})}\BibitemShut {NoStop}%
\bibitem [{\citenamefont {Chen}\ \emph {et~al.}(2003)\citenamefont {Chen},
  \citenamefont {Hida},\ and\ \citenamefont {Sanctuary}}]{full_phase}%
  \BibitemOpen
  \bibfield  {author} {\bibinfo {author} {\bibfnamefont {W.}~\bibnamefont
  {Chen}}, \bibinfo {author} {\bibfnamefont {K.}~\bibnamefont {Hida}}, \ and\
  \bibinfo {author} {\bibfnamefont {B.~C.}\ \bibnamefont {Sanctuary}},\ }\href
  {\doibase 10.1103/PhysRevB.67.104401} {\bibfield  {journal} {\bibinfo
  {journal} {Phys. Rev. B}\ }\textbf {\bibinfo {volume} {67}},\ \bibinfo
  {pages} {104401} (\bibinfo {year} {2003})}\BibitemShut {NoStop}%
\bibitem [{\citenamefont {Shaojin~Qin}(1997)}]{extrapol}%
  \BibitemOpen
  \bibfield  {author} {\bibinfo {author} {\bibfnamefont {L.~Y.}\ \bibnamefont
  {Shaojin~Qin}, \bibfnamefont {Yu-Liang~Liu}},\ }\href@noop {} {\bibfield
  {journal} {\bibinfo  {journal} {Phys. Rev. B}\ }\textbf {\bibinfo {volume}
  {55}},\ \bibinfo {pages} {2721} (\bibinfo {year} {1997})}\BibitemShut
  {NoStop}%
\end{thebibliography}%


%

\end{document}